    \documentclass[final,5p,times]{elsarticle}
\usepackage{amssymb}
\usepackage{amsmath}
  \usepackage{lineno}

\journal{Nuclear Physics B}

\begin{document}

\begin{frontmatter}



\title{Detectors and Electronics \\for the CBM experiment at~FAIR}


\author{Maksym Teklishyn for the CBM Collaboration} %% Author name

\affiliation{organization={GSI Helmholtzzentrum für Schwerionenforschung GmbH},%Department and Organization
            addressline={Planckstraße~1},
            city={Darmstadt},
            postcode={64291},
            country={Germany}}

%% Abstract
\begin{abstract}
The Compressed Baryonic Matter (CBM) experiment is a next-generation heavy-ion experiment under development at the future FAIR facility in Darmstadt, Germany. It is designed to explore the QCD phase diagram at high net-baryon densities with unprecedented precision. Operating in fixed-target mode with a continuous beam of up to $11\,A\mathrm{GeV}$ for heavy ions and $26\,\mathrm{GeV}$ for protons, CBM will investigate rare probes such as multi-strange hyperons, hypernuclei, and di-leptons, aiming to identify signatures of a first-order phase transition and the QCD critical point.

To achieve these goals, CBM employs a free-streaming, self-triggered readout architecture and a suite of radiation-hard, low-mass detectors capable of operating at interaction rates up to $10\,\mathrm{MHz}$. The experimental set-up consists of several detector subsystems optimised for precise vertexing, tracking, particle identification, and event reconstruction. These subsystems have undergone extensive prototyping and validation campaigns, with many components already tested and integrated into existing experiments such as STAR/RHIC, HADES/SIS18, and E16/J-PARC. These efforts culminated in the realisation of the mCBM test set-up at the SIS18 accelerator, where key systems were successfully commissioned under realistic beam conditions.

This contribution provides a concise overview of the current status of detector development, series production, and validation efforts through both simulations and measurements.
\end{abstract}

% \begin{highlights}
% \end{highlights}

%% Keywords

\end{frontmatter}

%   \linenumbers


\section{Introduction}
The Compressed Baryonic Matter (CBM) experiment, currently under development at the forthcoming Facility for Antiproton and Ion Research (FAIR) in Darmstadt, Germany, aims at exploring the properties of strongly interacting matter at high baryon densities \cite{Leifels2025}. CBM will study the Quantum Chromodynamics (QCD)  phase diagram  in a region that remains largely unexplored, where dense nuclear matter may undergo a first-order phase transition or exhibit critical phenomena \cite{Senger2024, Hoehne2024}.

\begin{figure}[h!] \centering
 \includegraphics[width=1\linewidth]{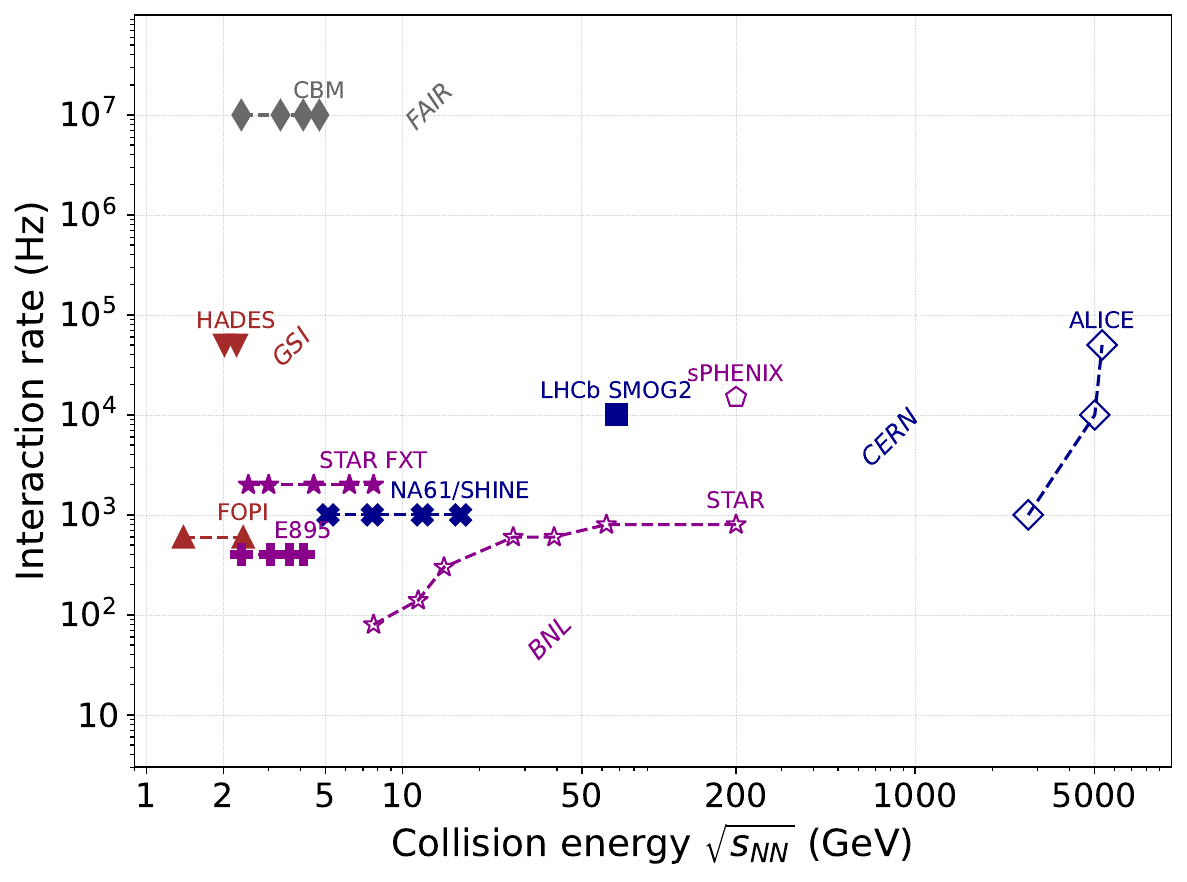}
\caption{%
Estimated interaction rates versus collision energy $\sqrt{s_{NN}}$ for heavy-ion experiments.
Filled markers indicate fixed-target experiments, open markers colliders.
The plot compares the projected performance of the future \textbf{CBM} experiment at \textbf{FAIR}
to existing and historical facilities at \textbf{GSI} (FOPI~\cite{Ritman_1995}, HADES~\cite{Harabasz_2023}),
\textbf{BNL} (E895~\cite{Rai_1999}, STAR~\cite{YANG2019951, Meehan_2016}, sPHENIX~\cite{Okawa_2023}),
and \textbf{CERN} (NA61/SHINE~\cite{Brzychczyk_2019}, ALICE~\cite{Valle_2025, Slupecki_2022, Triolo:2024lum}, fixed target LHCb runs~\cite{Garcia_2024, LHCb:2022qvj}).%
}
 \label{fig:cbm-rates}
\end{figure}

A central goal of the experiment is to probe the equation of state (EoS) and the phase structure of QCD matter through high-precision measurements of observables sensitive to the thermodynamic properties of the medium \cite{Agarwal2023}. These include the study of differential cross-sections and collective flow of multi-strange hadrons \cite{Lubynets2021, Khan:2021xyg, Golosov2022}, measurements of higher-order moments of conserved charges to detect critical fluctuations \cite{Vovchenko2020}, charmonium and open charm production \cite{Reichert:2025iwz, Sharma:2025ghp}, and analysis of di-lepton spectra as a diagnostic tool for the thermal history and possible phase transitions in the medium \cite{Jorge:2025wwp}.

To reach the statistical accuracy required for rare probe measurements, CBM will operate at unprecedented interaction rates of up to $10\,\mathrm{MHz}$. The experiment will employ fixed-target collisions using heavy-ion beams with kinetic energies between $2$ and $11\,A\mathrm{GeV}$, and proton beams up to $26\,\mathrm{GeV}$. For a comparison with existing facilities, see Fig.\,\ref{fig:cbm-rates}.

The ambitious physics program presents a unique set of challenges for the detector systems and readout infrastructure. The experiment must be able to precisely measure both low- and high-momentum tracks in an environment with high particle multiplicities and radiation levels. This requires detectors with a low material budget to reduce multiple scattering and secondary interactions. The detectors must also be fast, tolerant of radiation, and capable of sustaining high rates.

Given the use of continuous beams, precise timing information from the detectors is essential to disentangle overlapping events and to achieve reliable track-to-hit association. Instead of relying on a traditional hardware trigger, CBM will employ a novel free-streaming, self-triggered data acquisition scheme \cite{Kasinski2016}. In this architecture, data from all detector subsystems are continuously digitised and transmitted to a high-performance computing farm, where full online event reconstruction and selection are performed in real time \cite{Kisel2020}.

Furthermore, robust particle identification under conditions of high occupancy is crucial for the study of rare and short-lived particles such as hyperons and hypernuclei \cite{Vassiliev2017}. The CBM experiment will employ a suite of complementary detector systems to provide the necessary spatial, momentum, and timing resolution to fulfil these demanding requirements.

\begin{figure}[h!] \centering
 \includegraphics[width=.9\linewidth]{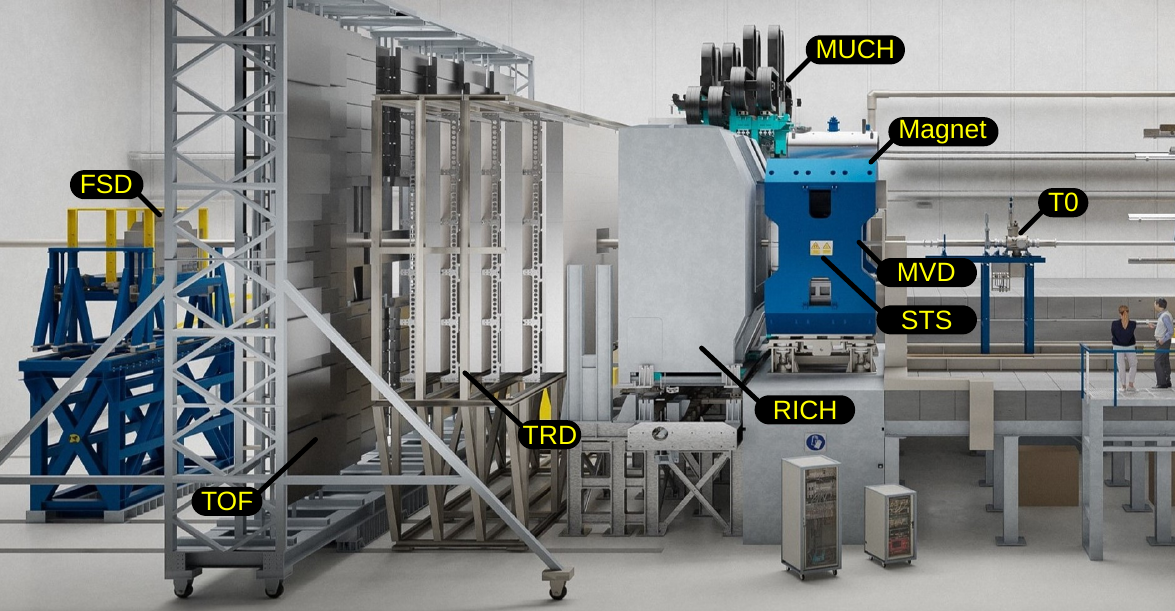}
 \caption{Schematics of the CBM experimental setup with its subdetectors (\textcopyright~ GSI/FAIR, Zeitrausch).}
 \label{fig:cbm}
\end{figure}

\section{Detector structure, tasks, and configurations}

The CBM detector system is designed to meet the demanding requirements of high-rate data acquisition and precise event reconstruction in a high-density environment; the structure of the experimental setup is depicted in Fig.\,\ref{fig:cbm}. The tracking and vertexing capabilities are provided by the Silicon Tracking System (STS) and the Micro-Vertex Detector (MVD). The STS, placed inside the $1\,\mathrm{Tm}$ superconductive dipole magnet, enables precise momentum measurement of charged particles. Upstream of the STS, the MVD reconstructs secondary decay vertices. Further details on their design and current status are provided in Sec.\,\ref{sec:track}.

Event geometry is determined using the Forward Spectator Detector (FSD), which registers spectator fragments emitted at small forward angles. This information is used to estimate the collision centrality and orientation of the reaction plane, which are essential for studying anisotropic flow and other observables sensitive to the initial geometry of the collision.

Global tracking and particle identification are accomplished through several dedicated detector systems. Electron identification is achieved with the Ring Imaging Cherenkov Detector (RICH) and the Transition Radiation Detector (TRD), covering complementary momentum ranges. The Time-of-Flight (ToF) wall, complimented by the T0 time reference detector, provides charged hadron identification based on precise timing measurements. In the forward region, the Muon Chamber (MUCH) system identifies muons for penetrating probes such as di-leptons. Together, these systems enable full event reconstruction and allow for the identification of rare signals in a high-occupancy environment.

% \begin{figure}[h!] \centering
%  \includegraphics[height=.3\linewidth]{ELEHAD.png} \hspace{.05\linewidth}
%  \includegraphics[height=.3\linewidth]{MUON.png} \hspace{.05\linewidth}
%  \includegraphics[height=.3\linewidth]{HADR.png}
%  \caption{Configurations of the CBM setup, left to right: ELEHAD, MUON, HADR.}
%  \label{fig:conf}
% \end{figure}

The CBM experimental setup is modular and can be flexibly reconfigured depending on the physics goals of a given measurement campaign.
% Three baseline configurations are shown in Fig.\,\ref{fig:conf}.
Each configuration includes a tailored set of detector subsystems, allowing the experiment to prioritise either lepton or hadron observables, or to reach maximum interaction rates for rare event studies.

The ELEHAD setup comprises the MVD, STS, RICH, TRD, ToF, and the FSD. This configuration is optimised for precision vertexing and tracking, with enhanced electron-to-pion separation. It is particularly suited for studying di-electron channels and strange hadrons at lower interaction rates, with a typical ``Day-1'' performance target of $0.1\,\mathrm{MHz}$.

The MUON setup includes the STS, Muon Chamber (MUCH), TRD, ToF, and FSD. The MVD and RICH are omitted in this case, while the MUCH system is introduced to enable efficient muon identification. This configuration is designed for measurements of low-mass and intermediate-mass di-muon pairs, such as thermal radiation and charmonia. The MUON setup can operate at average interaction rates of $1\,\mathrm{MHz}$ for initial running conditions.

The HADR configuration prioritises the highest possible interaction rates to access rare hadronic probes. It consists of the STS, TRD, ToF, and FSD, focusing on tracking and time-of-flight measurement while omitting the electron- and muon-specific detectors. This setup enables studies of multi-strange hyperons and hypernuclei with maximum luminosity, achieving up to $5-10\,\mathrm{MHz}$ in the mature phase.

\section{Tracking and vertexing detectors} \label{sec:track}
\subsection{Micro Vertex Detector}

The MVD is a key component of the CBM tracking system, dedicated to the precise reconstruction of secondary decay vertices and the suppression of combinatorial background in di-electron spectroscopy. It plays a crucial role in the identification of short-lived particles such as $\Lambda$, $\Xi$, and open charm hadrons, offering a vertex resolution on the order of $50\,\mu\mathrm{m}$. In addition, the MVD stations complement the STS by enabling the effective reconstruction of particles that decay with a neutral daughter, {\it e.g.}, $\Sigma$ baryons~\cite{Kisel2018}.

\begin{figure}[h!] \centering
 \includegraphics[width=.9\linewidth]{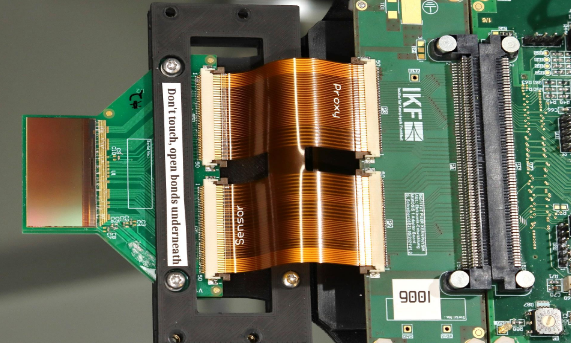}
 \caption{Test setup with two MIMOSIS-2.1 chips and the corresponding read-out electronics mounted at mCBM setup.}
 \label{fig:mvd}
\end{figure}

The MVD operates in vacuum and within the magnetic field environment of the CBM dipole magnet. The detector consists of four stations equipped with a total of approximately 300 radiation-tolerant CMOS pixel sensors. The core sensing element of the MVD is the MIMOSIS chip, a monolithic active pixel sensor specifically developed for the CBM experiment (see Fig.\,\ref{fig:mvd}). It is based on the well-established ALPIDE architecture and implemented in TowerJazz $180\,\mathrm{nm}$ CMOS technology. The sensor integrates a $27 \times 30\,\mu\mathrm{m}^2$ pixel cell with an in-pixel discriminator and features a fast, event-driven readout scheme with a typical readout time of $5\,\mu\mathrm{s}$ \cite{Klaus2019}.

Each chip hosts a pixel matrix of $1024 \times 504$ pixels
% , amounting to over half a million pixels per sensor,
and covers an area of approximately $4.2\,\mathrm{cm}^2$. The total sensor thickness is $50\,\mu\mathrm{m}$, with an epitaxial layer of around $25\,\mu\mathrm{m}$ and resistivity above $1\,\mathrm{k\Omega\cdot cm}$, ensuring both signal quality and radiation tolerance of $7 \times 10^{13}\,\mathrm{n_{eq}/cm^2}$ and $5\,\mathrm{Mrad}$ \cite{Darwish2023}. The sensor architecture incorporates multiple data concentration stages, an elastic output buffer, and eight switchable $320,\mathrm{Mbps}$ output links, enabling flexible and robust data handling for Au+Au interaction rates of up to $0.1\,\mathrm{MHz}$. Recent studies of the detector prototype during beam tests at the CERN SPS are presented in Ref.\,\cite{Deveaux2025}.

The power consumption ranges from $40$ to $70\,\mathrm{mW/cm^2}$ depending on the operational mode, which necessitates efficient thermal management within the MVD infrastructure. To ensure stable operation, an efficient thermal management system is implemented using CVD (Chemical Vapour Deposition) diamond or TPG (thermal pyrolytic graphite) carriers combined with actively cooled aluminium heat sinks \cite{Matejcek2024}.  The smallest functional unit of the detector is a quadrant, which integrates pixel sensors and thermal carriers to ensure mechanical stability and efficient heat evacuation \cite{Matejcek:2025fdh}.

\subsection{Silicon Tracking System}

The STS is the central tracking detector of the CBM experiment. It is responsible for precise momentum measurement of charged particles and contributes to vertex reconstruction. Positioned inside the dipole magnet, the STS operates in a high-multiplicity environment, handling hundreds of tracks per Au+Au collision. As such, it is designed to work in a self-triggered, free-streaming data acquisition mode, providing continuous tracking information.

The STS comprises 876 double-sided, double-metal (DSDM) silicon micro-strip modules arranged into eight tracking stations. These are divided into three upstream and five downstream stations, covering a solid angle of approximately $2.5^\circ$ to $25^\circ$. The sensors, produced by Hamamatsu Photonics \cite{hpk}, are $320\,\mu\mathrm{m}$ thick and available in various lengths, up to $12.4\,\mathrm{cm}$, with a common strip pitch of $58\,\mu\mathrm{m}$, with a $7.5^\circ$ stereo angle, enable two-dimensional position reconstruction.

Signal transmission is realised through  lightweight aluminium-polyimide microcables, allowing the readout electronics to be located outside the detector acceptance. Each micro-strip module connects to a pair of Front-End Boards (FEBs), hosting eight custom SMX (STS-MUCH-XYTER) ASICs for analogue signal processing \cite{Kasinski2018}. These chips enable simultaneous time and amplitude measurement with 5-bit flash ADCs and 14-bit time-to-digital converters, achieving time resolutions on the order of $5\,\mathrm{ns}$. Together, the sensors and readout form a five-dimensional tracking system capable of measuring position, time, and deposited energy \cite{Teklishyn2024}.

Mechanical integration is achieved using lightweight carbon-fibre ladders and aluminium support structures, assembled into modular Half-Units. The material budget per station ranges from $0.3\%$ to $1.4\%\,X_0$, ensuring minimal impact from multiple scattering. The detector features a low dead time per channel of approximately $300\,\mathrm{ns}$, enabling reliable operation under high-occupancy conditions~\cite{Teklishyn2025}.

Ongoing series production and quality assurance efforts have validated the design, with more than one-third of the modules assembled and characterised \cite{Rodr_guez_Rodr_guez_2025}. The STS has also been commissioned in beam conditions as part of the mCBM test setup at SIS18 \cite{CBM:2025voh}, and selected modules are deployed in the E16 experiment at J-PARC \cite{Aoki_2025}. Its design and performance serve as a technological reference for other high-rate tracking systems in nuclear and medical applications \cite{DPG2025}.

\section{Event geometry determination} \label{sec:geo}

\subsection{Forward Spectator Detector}

The FSD is one of the dedicated subsystems of the CBM experiment, designed to measure the collective behaviour of nuclear matter, with a particular focus on anisotropic flow. It supersedes the earlier development of the Projectile Spectator Detector. Positioned at forward rapidity, the FSD detects spectator nucleons and fragments originating from the nuclear remnants after a collision at the target. Its primary role is to provide accurate event plane reconstruction and collision centrality determination at interaction rates up to $10\,\mathrm{MHz}$.

The FSD is currently designed as a scintillator hodoscope consisting of charge-sensitive scintillator tiles. The segmentation pattern is optimised according to the expected particle flux, with finest granularity of $4 \times 4\,\mathrm{cm}^2$ near the beam pipe where the particle density is highest, and coarser granularity at larger radii.
% This ensures efficient detection and identification of spectator fragments over a wide range of impact parameters.

The FSD aims to distinguish between fragments with $Z=1$ and $Z>1$. Event plane reconstruction is performed based on the azimuthal asymmetry of spectator distributions, employing the $Q$-vector method  \cite{Poskanzer1998} and multiple subevent techniques to enhance resolution. Simulation studies have demonstrated that the FSD achieves an event plane resolution of approximately 70\% for the $x$-component and about 40\% for the $y$-component in Au+Au collisions at $11\,A\mathrm{GeV}$ \cite{Dvorak2025}.

\section{Global tracking and particle identification} \label{sec:pid}

\subsection{Beam Monitoring and Time-Zero System } \label{sec:bmon}

The Beam Monitoring and Time-Zero System (BMON) system of the CBM experiment provides the start time for time-of-flight measurements and monitors the beam halo as part of the fast Beam Abort System (BAS). It consists of two detector stations, T0 and Halo, mounted in the vacuum chambers upstream of the target~\cite{Rost2023}.

The T0 station employs a single $70\,\mu\mathrm{m}$ thick poly crystalline CVD diamond sensor with $1\times1\,\mathrm{cm^2}$ active area and $16\times16$ orthogonal strip metallisation. It delivers position and timing information with $\leq 50\,\mathrm{ps}$ resolution and readout rates up to $1\,\mathrm{MHz}$ per channel. Amplified and digitised using PADI~\cite{Ciobanu2014} and GET4~\cite{Deppe2009} ASICs, data is transmitted to the CBM Data Acquisition (DAQ) via a Common Readout Interface (CRI) interface \cite{10.1117/12.2501415}.

\begin{figure}[h!] \centering
 \includegraphics[width=.9\linewidth]{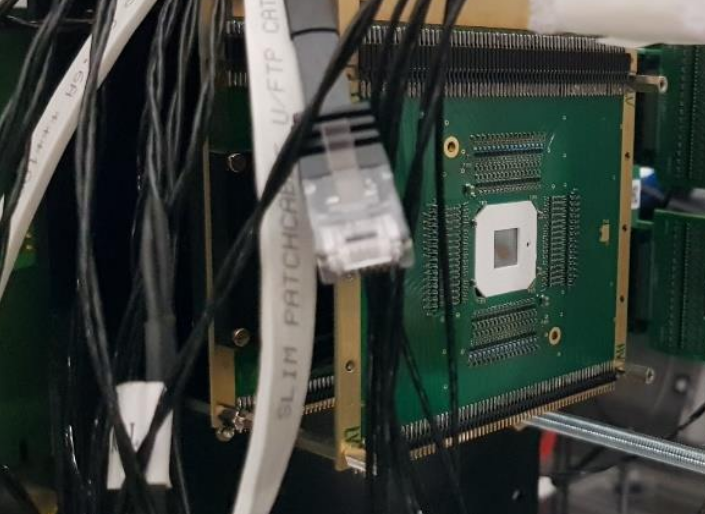}
 \caption{LGAD sensor mounted on the AC-coupled front end board.}
 \label{fig:lgad}
\end{figure}

The Halo station comprises four smaller diamond sensors in a mosaic layout, enabling beam edge monitoring and integration with the BAS logic. Both sensors are mounted on a retractable manipulator for safe operation under high-intensity beams. LGAD technology is under evaluation as an alternative for proton beams~\cite{Krueger2022}. The LGAD sensor test setup is shown in Fig.\,\ref{fig:lgad}.

Simulations show negligible impact of the T0 sensor on STS performance, confirming its compatibility with high-precision tracking. The BMON system ensures reliable timing, beam quality monitoring, and fast beam abort capabilities essential for CBM operation at $\leqslant10^7\,\mathrm{Hz}$.

\subsection{Ring Imaging Cherenkov system}

The Ring Imaging Cherenkov system is designed for electron identification at momenta up to $8\,\mathrm{GeV}/c$; it achieves an overall pion suppression factor between 1000 and 5000.

The RICH follows a robust design optimised for the high interaction rates (up to $10\,\mathrm{MHz}$) and high particle multiplicities encountered in central Au+Au collisions. It uses a $1.7\,\mathrm{m}$ long $\mathrm{CO}_2$ radiator volume operated at a slight overpressure of $2\,\mathrm{mbar}$, with a pion Cherenkov threshold of about $4.65\,\mathrm{GeV}/c$. Cherenkov photons are reflected by a system of lightweight spherical mirrors made from SIMAX glass, coated with Al+MgF$_2$, ensuring high UV reflectivity. The mirrors, arranged into two planes above and below the beam pipe, have a curvature radius of $3\,\mathrm{m}$ and a thickness of $6\,\mathrm{mm}$.

Photon detection is performed by an array of Multi-Anode PhotoMultiplier Tubes (MAPMTs) of type H12700 from Hamamatsu. The system consists of approximately 1100 MAPMTs, covering a total area of about $2.4\,\mathrm{m}^2$, and includes around 55,000 channels. The MAPMTs are equipped with p-terphenyl wavelength-shifting (WLS) films to enhance their sensitivity to UV photons. Without WLS films, about 24 hits per electron ring are recorded, with an improvement of approximately 20\% when WLS is applied. A shielding box and mirror tilting scheme are implemented to reduce the effect of the dipole stray magnetic field to below $1-2\,\mathrm{mT}$ at the photocathode plane \cite{AdamczewskiMusch2017, AdamczewskiMusch2020}.

The mechanical design of the mirror holding structure has been optimised to minimise the material budget while ensuring mechanical stability. Individual mirror tiles are mounted with three-point fixation, allowing for precision alignment and maintaining the surface quality necessary for sharp ring imaging \cite{Becker2024}.

\subsection{Transition Radiation Detector}

The TRD in CBM is designed for electron identification and light fragment detection in a high-rate heavy-ion collision environment. It complements the RICH and ToF systems to achieve a high pion suppression factor and also contributes essential information for global tracking.

The CBM TRD is based on four layers of multi-wire proportional chambers (MWPCs) equipped with polyethylene (PE) foam radiators and operated with a Xe/CO$_2$ (85:15) gas mixture. Electrons traversing the radiators emit transition radiation (TR) photons, which are then absorbed and detected in the MWPCs, in addition to their ionisation energy loss. Electron identification is achieved by detecting this excess energy deposition, which distinguishes electrons from hadrons that typically do not produce significant TR.

The MWPCs feature a thin $5\,\mathrm{mm}$ drift region and symmetric $3.5\,\mathrm{mm}$ amplification gaps, ensuring fast signal collection with drift times between $250$ and $300\,\mathrm{ns}$. The entrance windows of the chambers are made from aluminised polyimide foils, stabilised by a carbon grid to limit window deformation under operational pressure.

Signal readout is performed through cathode-pad planes, achieving a position resolution of approximately $300,\mu\mathrm{m}$. The analogue signals are processed by SPADIC front-end ASICs (Self-triggered Pulse Amplification and Digitisation integrated Circuit), featuring a fully digitised, self-triggered architecture with continuous ADC sampling at $160\,\mathrm{MHz}$ \cite{Armbruster:2010qwa}. Each SPADIC chip contains 32 channels, each with a charge-sensitive preamplifier, a shaper, and a pipeline of 32 time-sampled ADC values per event. The chip also supports event time-stamping and implements internal zero-suppression, which significantly reduces the data volume for transmission.

In addition to signal digitisation, SPADIC provides pulse-shape information useful for cluster centroiding and noise discrimination \cite{Schledt_2023}. Its design enables dead-time-free operation in free-streaming mode and allows for precise timing with sub-nanosecond resolution. The self-triggered nature of the chip makes it fully compatible with CBM's triggerless DAQ architecture.

Beam tests and laboratory measurements demonstrated an electron detection efficiency of about $(98.5 \pm 2.0)\%$, with energy resolutions matching the CBM performance requirements. Stability tests at the CERN Gamma Irradiation Facility have confirmed the chamber's capability to operate under high radiation loads without significant gain degradation or space-charge effects.

Simulation studies with Au+Au collisions at $10\,A\mathrm{GeV}$ predict trigger rates up to several hundred $\mathrm{kHz}$ per module. To cope with these rates, the TRD employs cathode pads of variable size (ranging from $1.2\,\mathrm{cm}$ to $8.0\,\mathrm{cm}$) and rotated module orientations in alternating layers for optimal 2D tracking. Besides its primary role in particle identification, the TRD also offers important input to global tracking \cite{Kaehler2020}.

To further enhance spatial resolution and low-$p_T$ tracking, a two-dimensional (2D) MWPC readout concept, {TRD-2D}, has been proposed. This scheme combines induced-charge readout on cathode pads for the $x$-coordinate with direct anode-wire signals for the $y$-coordinate, enabling simultaneous two-dimensional localisation of each cluster. Compared to the standard TRD, which provides a spatial resolution of approximately $300\,\mu\mathrm{m}$ in $x$ and a coarse $\sim\!5\,\mathrm{mm}$ in $y$, the TRD-2D achieves about $150\,\mu\mathrm{m}$ in $x$ and $850\,\mu\mathrm{m}$ in $y$, thus delivering sub-millimetre tracking capability in both directions. Importantly, its performance remains robust at design particle fluxes of up to $30\,\mathrm{kHz/cm}^2$, corresponding to $10\,\mathrm{MHz}$ central Au+Au collision rates, without degradation in rate capability~\cite{Bercuci2021}.

\subsection{Time-of-Flight Detector}

The Time-of-Flight (ToF) system of the CBM experiment provides charged hadron identification (pions, kaons, and protons) over a momentum range up to approximately $4\,\mathrm{GeV}/c$ within a polar angular coverage of $2.5^\circ$ to $25^\circ$. Designed for continuous high-rate operation at up to $10\,\mathrm{MHz}$ interaction rates, the ToF wall must combine fast timing, high efficiency, and resilience to intense particle fluxes.

The ToF detector consists of a large wall with an active area of about $120\,\mathrm{m}^2$, incorporating around 230 modules and 1400 Multi-gap Resistive Plate Chambers (MRPCs), resulting in approximately $90{,}000$ readout channels. The design is optimised for different occupancy regions:
\begin{itemize}
    \item \textbf{High-rate region} (up to $50\,\mathrm{kHz/cm}^2$): low-resistivity glass MRPCs (MRPC1 type) with small granularity ($6{-}20\,\mathrm{cm}^2$ per strip).
    \item \textbf{Intermediate-rate region} ($1{-}5\,\mathrm{kHz/cm}^2$): MRPC3a detectors with strip granularity ($1\times27\,\mathrm{cm}^2$).
    \item \textbf{Low-rate region} ($<1\,\mathrm{kHz/cm}^2$): MRPC3b and MRPC4 detectors using commercial float glass, with larger strip lengths up to $50\,\mathrm{cm}$.
\end{itemize}

The system aims for a full system time resolution better than $80\,\mathrm{ps}$ and an overall detection efficiency exceeding $95\%$. Occupancy across the wall is kept below $5\%$, even at maximum rates. The MRPC types are tuned to achieve the necessary rate capability by selecting appropriate electrode materials: low-resistivity glass ($\rho \sim 10^{10}\,\Omega\,\mathrm{cm}$) for high-rate regions, and thin float glass for lower-rate areas.

Each MRPC channel is read out via a front-end electronics chain consisting of PADI (Preamplifier-Discriminator) ASICs~\cite{Ciobanu2014}, which convert the analogue MRPC signals into digital pulses with sub-nanosecond jitter. These signals are then time-stamped with the GET4 TDC ASIC~\cite{Deppe2009}, which provides a time binning of $25\,\mathrm{ps}$ and can handle high hit rates with low dead time. The digitised data are transmitted via optical links to the CBM Data Acquisition System using the CRI~\cite{10.1117/12.2501415}, which also provides slow control, monitoring, and clock distribution. The entire electronics chain supports free-streaming readout and time-based event reconstruction within the First-Level Event Selector (FLES) environment.

Prototypes of each MRPC type have been extensively tested. In-beam measurements and cosmic ray setups confirmed a single-counter time resolution of around $50\,\mathrm{ps}$ and high efficiency under free-streaming readout conditions. Importantly, noise rates are acceptable even in the high-rate regions, with design margins validated through long-term tests~\cite{Deppner2019}.

A subset of CBM ToF detectors and readout electronics was deployed as part of the {eTOF} (endcap Time-of-Flight) system in the STAR experiment at RHIC. Installed during the Beam Energy Scan II campaign, the eTOF extended STAR’s particle identification capabilities to forward rapidities and served as a large-scale integration and performance test for CBM ToF technology \cite{STAR:2016gpu}. The system included MRPC3b counters with CBM-specific front-end electronics and readout, successfully operating under RHIC conditions with high stability and excellent timing performance. The experience gained from eTOF was instrumental in validating the design choices of the CBM ToF wall, including detector construction, electronics reliability, and integration with a large-scale DAQ infrastructure \cite{Weidenkaff2023}.

\subsection{Muon Chamber}

The Muon Chamber (MuCh) is a dedicated subsystem of the CBM experiment, designed for the identification of muons originating from the decay of vector mesons and charmonia. It enables the study of penetrating probes that carry information from the early stages of heavy-ion collisions, even in the presence of high-density nuclear matter.

The MuCh system consists of multiple detector stations interleaved with segmented hadron absorbers. It covers a polar angular range of approximately $5^\circ$ to $25^\circ$ and is optimised for efficient operation at high interaction rates, up to $10\,\mathrm{MHz}$. The absorber configuration includes alternating layers of iron and carbon-based composites to suppress hadrons while allowing muons to pass through to the detector planes.

The tracking stations are based on Gas Electron Multiplier (GEM) and Resistive Plate Chamber (RPC) technologies \cite{Kumar:2021xpj}. The first two stations employ triple-GEM detectors for their excellent spatial resolution and rate capability. The final two stations are equipped with RPCs optimised for lower particle fluxes. The readout plane in each GEM module follows a progressive pad geometry, with the inner zone featuring high granularity ($0.1\,\mathrm{cm}^2$ pads) and outer regions using pads up to $2.89\,\mathrm{cm}^2$ to match the radial intensity profile of the CBM acceptance.

Signal processing in the MuCh detectors is performed using the {SMX}  front-end ASIC~\cite{Kasinski2018}, a custom-designed, self-triggered chip developed jointly for the STS and MuCh subsystems. The versatile design allows it to operate in multiple amplification modes, making it suitable for different detector technologies. In the MuCh configuration, the chip is typically used in low-gain mode, providing a dynamic range below $100\,\mathrm{fC}$ to match the expected charge signals from GEM and RPC detectors.

Extensive performance studies were conducted during mCBM test campaigns at GSI. GEM modules were operated under realistic beam conditions, using $^{69}\mathrm{Au}$ or $^{73}\mathrm{U}$ beams at energies up to $1.23\,A\mathrm{GeV}$ and instantaneous interaction rates up to $2.5 \times 10^8$ per 10-second spill. The setup used up to 18 FEBs per module with a full DAQ chain and online monitoring. Digi rates of up to $200\,\mathrm{kHz/cm^2}$ were observed in the inner zone, confirming the rate capability for the final CBM configuration. Spatial and timing correlations between GEM planes demonstrated precise alignment and stability, with typical time offsets below $13\,\mathrm{ns}$ and spatial resolution preserved across modules \cite{Ghosh:2024ofo, Ghosh:2025krh}.

Simulation studies and prototype measurements have confirmed that the MuCh design satisfies the requirements for efficient di-muon reconstruction, with sufficient granularity, rate handling, and radiation tolerance to ensure robust operation during extended high-luminosity physics runs.

\section{Data Acquisition and Online Event Processing}

The CBM experiment employs a novel, triggerless DAQ architecture designed to handle continuous data streams from all detector subsystems at interaction rates up to $10\,\mathrm{MHz}$. The system is fully self-triggered, with front-end electronics digitising and transmitting data as soon as a detector signal is registered. This architecture eliminates latency and bandwidth limitations associated with traditional hardware triggers and is essential for rare-probe physics in high-rate heavy-ion collisions.

\subsection{Front-End and Readout Electronics}

Each detector subsystem is equipped with dedicated front-end ASICs optimised for fast, low-noise operation in a free-streaming mode. These include the SMX chip for STS and MuCh~\cite{Kasinski2018}, GET4 for fast timing detectors such as ToF and BMON~\cite{Deppe2009}, and SPADIC for the TRD~\cite{Armbruster:2010qwa}. These ASICs digitise the time and amplitude of incoming signals and apply local zero suppression to reduce data volume.

Digitised data are transmitted via high-speed serial links (GBT protocol) to detector-specific Readout Boards (ROBs). The ROBs perform data aggregation and forward the event fragments via optical fibres to a CRI. The CRI handles slow control, clock and trigger distribution \cite{Sidorenko:2021hmw}, and data formatting, and interfaces directly with the computing farm over Gigabit Ethernet or optical links \cite{Zabolotny:2017hwz}.

All detector data are time-stamped with a global clock and aligned by their timestamps during online event building.

\subsection{First-Level Event Selector}

At the core of CBM’s online processing is the {First-Level Event Selector (FLES)}, a high-performance computing cluster responsible for real-time event building, reconstruction, and filtering. The FLES receives continuous data streams from all detectors, sorts the data by global timestamps, and builds physical events within a programmable time window \cite{Akishina:2015mfa, Kisel:2020lpa}.

FLES nodes are equipped with multi-core CPUs, large memory buffers, and optionally GPUs or FPGAs for parallelised processing. The software stack includes fast algorithms for global tracking, vertexing, time-based clustering, and online calibration. Advanced filtering algorithms are employed to select events of interest, reducing the output data rate by several orders of magnitude before persistent storage.

The performance of the FLES has been validated using prototype installations and simulations under full-load conditions. Time-based event building and processing have been demonstrated for interaction rates exceeding $10\,\mathrm{MHz}$, with reconstruction latencies on the order of milliseconds. Data throughput tests have shown that the system can sustain up to 1~TByte/s input bandwidth across subsystems.

\subsection{System integration and Phase-0 deployment}

The full readout and FLES system has been deployed and tested as part of the mCBM programme at SIS18. Detector subsystems including STS, TRD, MuCh, ToF, BMON and others were successfully operated in free-streaming mode using the final DAQ infrastructure. Realistic beam conditions confirmed the stability and robustness of the data flow from front-end to online event processing.

The CBM DAQ and FLES architecture represents a paradigm shift in heavy-ion experiment design, enabling the experiment to operate at unprecedented interaction rates while maintaining the flexibility and precision needed for rare-probe physics.

\section{Conclusion}

The detector systems have been carefully designed to address the extreme demands imposed by the physics goals: high granularity, fast response, radiation hardness, and low material budget. Extensive prototyping, beam testing, and integration into existing experiments such as STAR, HADES, and E16 have provided critical validation of the chosen technologies.

With series production and commissioning efforts well underway, the CBM collaboration has realised a modular and reconfigurable experimental setup capable of precise tracking, vertexing, and particle identification. The readiness of subsystems demonstrates the maturity of the detector concept and its implementation.

As the FAIR facility approaches completion, CBM is uniquely positioned to deliver long-awaiting results on the QCD matter phase structure, critical phenomena, and the equation of state of dense baryonic matter. The developments reported in this work mark a decisive step toward realising these scientific objectives.

\bibliographystyle{elsarticle-num}
\bibliography{trimmed_authors}

\end{document}